\documentclass[aps,prl,twocolumn,superscriptaddress,showpacs]{revtex4}

\usepackage{amsmath}
\begin{document}

%\preprint{}

\title{Spin continuity equation and definition of spin current}


\author{Xiang Zhou}
\email[To whom correspondence should be addressed:~]{xzhou@mac.com}
%\thanks{}
\affiliation{Department of Physics, Wuhan University, Wuhan 430072, China}
\affiliation{Department of Physics, Texas A\&M University, College Station, Texas 77843, USA}
\author{Zhenyu Zhang}
\affiliation{Department of Physics, Wuhan University, Wuhan 430072, China}
\author{Cheng-Zheng Hu}
\affiliation{Department of Physics, Wuhan University, Wuhan 430072, China}


\date{\today}

\begin{abstract}
For an electron, a spin-$1/2$ particle, the spin charge $\mathbf{s}$, a real pseudovector with constant length, could determine the spin polarization properties in quantum mechanics. Since spin density $\rho_{\mathbf{s}}$ could be expressed as the product of probability density $\rho$ and the spin charge $\mathbf{s}$, the spin continuity equation could be derived from fundamental principles of quantum mechanics and the definition of spin current arise naturally. Only in some specific conditions the conventional definition of spin current coincides with ours. The equilibrium spin currents would vanish automatically in two-dimensional electron gas with Rashba spin-orbit interaction by using our definition of spin current.
\end{abstract}

%\keywords{}

\maketitle

%\section{}
%\subsection{}
%\subsubsection{}

Spin current is one of the most important physical quantities in spintronics. However, there is still not an unambiguous definition of spin current\cite{Rashba2003,Rashba2007}.  New definitions of spin current had been proposed \cite{Sun2005,Wang2006,Zhang2006,Jin2006} and  explanations of the conventional definition of spin current had been made \cite{Sonin2007a,Sonin2007b,Tokatly2008}. Although the interests and publications of research in spin dynamics and transport still increased in spite of the debates on the definition of spin current, the ambiguity of definition of spin current should and must be eliminated as soon as possible. In the general transport theory the current density of a local charge is defined from the continuity equation. It was surprising that \emph{spin continuity equation} which is the continuity equation of spin transport could not be easily obtained by computing the time derivative of spin density. The reasons come from that the spin continuity equation would contain the extra torque terms because of the non-conservation of spin and the algebraic properties of the spin operator $\hat{\mathbf{s}}=\frac{\hbar}{2}\hat{\boldsymbol{\sigma}}$ is complex, where $\hat{\boldsymbol{\sigma}}$ are the Pauli matrices. In quantum mechanics the state of an electron, a spin-$1/2$ particle, is described by the spinor $\Psi(\mathbf{r},t)$ and $\rho(\mathbf{r},t)=\Psi^\dagger(\mathbf{r},t)\Psi(\mathbf{r},t)$ is the probability density. The spin density $\rho_{\mathbf{s}}=\Psi^\dagger(\mathbf{r},t)\hat{\mathbf{s}}\Psi(\mathbf{r},t)$ had been well defined \cite{Pauli1973}. Comparing to the electronic charge density $\rho_e=\Psi^\dagger(\mathbf{r},t)e\Psi(\mathbf{r},t)$, one might regard the spin operator $\hat{\mathbf{s}}$ as the \emph{spin charge} which is the local charge in spin continuity equation. However, 
using the general form [charge density]=[charge]$\times$[density], one might intuitionally define the spin charge
\begin{eqnarray}
\mathbf{s}(\mathbf{r},t) & := & \frac{\rho_{\mathbf{s}}(\mathbf{r},t)}{\rho(\mathbf{r},t)}=\frac{\Psi^\dagger(\mathbf{r},t)\hat{\mathbf{s}}\Psi(\mathbf{r},t)}{\Psi^\dagger(\mathbf{r},t)\Psi(\mathbf{r},t)}  \nonumber \\
& = & \chi^\dagger(\mathbf{r},t)\hat{\mathbf{s}}\chi(\mathbf{r},t)=\frac{\hbar}{2}\mathbf{n}(\mathbf{r},t),
\end{eqnarray}
where the $\mathbf{n}(\mathbf{r},t)=\chi^\dagger(\mathbf{r},t)\hat{\boldsymbol{\sigma}}\chi(\mathbf{r},t)$ is the polarization vector that corresponds to the normalized spinor $\chi(\mathbf{r},t)=\rho^{-\frac{1}{2}}(\mathbf{r},t)\Psi(\mathbf{r},t)$ \cite{Schiff1968}.

In this letter, we show that the spin charge $\mathbf{s}$ could determine the spin polarization properties of an electron. The spin continuity equation would be easily derived when the spin density is expressed as $\rho_{\mathbf{s}}=\rho\mathbf{s}$ since the spin charge $\mathbf{s}$ is an real pseudovector with constant length.  Then the definition of spin current density would arise naturally. The equilibrium spin currents in two-dimensional (2D) electron gas with Rashba spin-orbit interaction (SOI) would vanish automatically by using our definition of spin current.

It is well known that the polarization vector $\mathbf{n}$  could totally determine the density matrix $\hat{\varrho}=\frac{1}{2}(\mathit{1}+\mathbf{n}\cdot\hat{\boldsymbol{\sigma}})$ of an electron in a pure normalized spinor  \cite{Fano1957,Blum1996}. The polarization vector $\mathbf{n}$ was introduced in nuclear magnetic resonance (NMR) by Bloch to simplify the treatment of spin precession \cite{Bloch1946,Wangsness1953}. The polarization vector $\mathbf{n}$ of an electron is a unit real pseudovector in three-dimensional space \cite{Robson1974} and could be measured by experiments \cite{Kessler1985}. In other words, the spin charge $\mathbf{s}=\frac{\hbar}{2}\mathbf{n}$, a real pseudovector with constant length, could totally determine the spin polarization properties of an electron \cite{Landau1998}. Moreover, the pseudovector $\mathbf{s}$ and the spin density $\rho_{\mathbf{s}}=\rho\mathbf{s}$ could be sophistically obtained by using the mathematical language \emph{Geometric Algebra} where the vector $\mathbf{s}$ is called spin vector \cite{Hestenes1971,Hestenes1991,Hestenes2003,Baylis1996,Doran2003}.  We rename $\mathbf{s}$ as \emph{spin charge} to emphasize that it functions as the charge in the spin continuity equation. 

Since the spin density could be expressed as $\rho_{\mathbf{s}}=\rho \mathbf{s}$ and the spin charge $\mathbf{s}$ is a real pseudovector with constant length, the spin continuity equation could be easily derived from the time derivative of the spin density. Below we derive the spin continuity equation by an alternative method. In quantum mechanics, the fundamental probability continuity equation is
\begin{equation}
\frac{\partial\rho}{\partial t}+\nabla\cdot\mathbf{j}=0,
\label{pce}
\end{equation}
where $\mathbf{j}$ is the probability current density. We multiply the spin charge $\mathbf{s}$ to both sides of Eq.~(\ref{pce})
\begin{equation}
\mathbf{s}(\frac{\partial\rho}{\partial t}+\nabla\cdot\mathbf{j})=0.
\end{equation}
After moving the $\mathbf{s}$ into the time derivative and divergence operators, one could obtain the spin  continuity equation
\begin{equation}
\frac{\partial(\rho\mathbf{s})}{\partial t}+\nabla\cdot(\mathbf{j}\mathbf{s})=\rho\frac{\partial\mathbf{s}}{\partial t}+\mathbf{j}\cdot\nabla\mathbf{s},
\label{sce}
\end{equation}
where the $\mathbf{j}\mathbf{s}$ is a dyadic tensor and the rule of the divergence of a dyadic tensor is $\nabla\cdot(\mathbf{A}\mathbf{B})=\mathbf{B}\nabla\cdot\mathbf{A}+\mathbf{A}\cdot\nabla\mathbf{B}$. Our derivation of the spin continuity equation is based on the fundamental principles of quantum mechanics. It is thus reasonable to define the spin current density
\begin{equation}
\mathcal{J}_{\mathbf{s}}:=\mathbf{j}\mathbf{s}.
\label{mscd}
\end{equation}
The spin current density $\mathcal{J}_{\mathbf{s}}$ is a \emph{measurable} 2-rank real pseudotensor. Our definition of spin current density is similar to the definition of spin current density in a modified phenomenological Bloch equation \cite{Ziese2001}.

The two r.h.s. terms in Eq.~(\ref{sce}) describe the torques. The first torque term is just the product of probability density $\rho$ and classical torque $\partial\mathbf{s}/\partial t$. The second torque term whose physical meaning might not be obvious is the dot product of the gradient of the spin charge $\nabla\mathbf{s}$ and the probability current density $\mathbf{j}$. Considering an electron in a magnetic field $\mathbf{H}$, the Hamiltonian is $-\gamma\frac{1}{2}\hat{\boldsymbol{\sigma}}\cdot\mathbf{H}$ where $\gamma$ is the gyromagnetic ratio. The spin precession satisfies the classical equation \cite{Bloch1946,Wangsness1953}
\begin{equation}
\frac{\partial \mathbf{s}}{\partial t}=-\gamma\mathbf{H}\times\mathbf{s}.
\end{equation}
If the magnetic field $\mathbf{H}(\mathbf{r})$ is time-independent, the spin charge vector $\mathbf{s}(\mathbf{r})$ of the normalized eigenstate is also time-independent. The spin charge $\mathbf{s}(\mathbf{r})$  is then parallel or anti-parallel to the magnetic field $\mathbf{H}(\mathbf{r})$ and the torque ${\partial \mathbf{s}}/{\partial t}$ vanishes. At that time only the second torque term could describe the torque acts on the electron.

Conventionally, the probability current density $\mathbf{j}$ could be expressed as $\mathbf{j}=\mathrm{Re}\Psi^\dagger\hat{\mathbf{v}}\Psi$, where $\hat{\mathbf{v}}=\partial \hat{H}/\partial\hat{\mathbf{p}}$ and $\hat{H}$ is the Hamiltonian. The $i$-direction $k$-component  spin current density is
\begin{equation}
\mathcal{J}_{\mathbf{s}}^{ik}=j_is_k=\mathrm{Re}\Psi^\dagger v_i\Psi\cdot\Psi^\dagger\hat{s}_k\Psi/\Psi^\dagger\Psi.
\end{equation}
When the state $\Psi$ is the eigenstate of either $\hat{v}_i$ or $\hat{s}_k$, it is easily to prove that 
\begin{equation}
\mathcal{J}_{\mathbf{s}}^{ik}=\mathrm{Re}\Psi^\dagger \frac{1}{2}\{\hat{v}_i,\hat{s}_k\}\Psi,
\label{cscd}
\end{equation}
which has the same form as the \emph{conventional} definition of spin current density\cite{Rashba2003}.

It was noticed by Rashba that there would be equilibrium spin currents in the 2D electron gas with Rashba SOI by using the conventional definition of spin current \cite{Rashba2003} and debates have arisen from then on \cite{Sun2005,Wang2006,Zhang2006,Jin2006,Sonin2007a,Sonin2007b,Tokatly2008}. Since the state $\Psi$ is usually not the eigenstate of either $\hat{v}_i$ or $\hat{s}_k$, the $i$-direction $k$-component  spin current density $\mathcal{J}_{\mathbf{s}}^{ik}$ could not be expressed in the conventional form Eq.~(\ref{cscd}).  We can show that the equilibrium spin currents would vanish automatically by using our definition of spin current. In the 2D electron gas with Rashba SOI the spin polarization is perpendicular to the $\mathbf{k}$ in the eigenstates $\psi_{\lambda}(\mathbf{k})$, where $\lambda=\pm$ correspond to the upper and lower branches of the spectrum, respectively \cite{Rashba2003}. In any eigenstate the spin charge $\mathbf{s}$ is conserved, the spin current density is divergenceless and the two torque terms vanish. For an arbitrary eigenstate $\psi_{+}(\mathbf{k})$ in upper branch, there always exists a corresponding eigenstate in lower branch $\psi_{-}(\mathbf{k'})$ which has the same velocity $\mathbf{v}$. The spin current densities in the two eigenstates $\psi_{+}(\mathbf{k})$ and $\psi_{-}(\mathbf{k'})$ are opposite to each other since their probability current densities $\mathbf{j}_+$ and $\mathbf{j}_-$ are equal but their spin polarizations $\mathbf{s}_+$ and $\mathbf{s}_-$ are opposite. It would be easily proved that the energy of the two eigenstates $\psi_{+}(\mathbf{k})$ and $\psi_{-}(\mathbf{k'})$ are the same. According to the principles of statical physics, the two eigenstates $\psi_{+}(\mathbf{k})$ and $\psi_{-}(\mathbf{k'})$ have the same statistical probability in thermodynamic equilibrium. The contribution in total spin current density provided by the two spin current density in $\psi_{+}(\mathbf{k})$ and $\psi_{-}(\mathbf{k'})$ would be zero. Thus in thermodynamic equilibrium the total spin current in 2D electron gas with Rashba SOI would vanish.

In conclusion, the spin continuity equation of an electron has been derived from the time derivative of the spin probability. The spin current density has been defined naturally from the spin continuity equation. The relationship between our definition and the conventional definition of spin current is discussed. The equilibrium spin currents would vanish automatically in 2D electron gas with Rashba SOI by using our definition of spin current.

X. Z. gratefully acknowledge discussions with Z. K. Meng and X. J. Liu and is supported by CSC. This work is supported by NSFC (No. 10534030).

\end{document}